\documentclass[sn-nature]{sn-jnl}%

\RequirePackage{amsmath,amsfonts,amssymb}
\RequirePackage{mathptmx}
\usepackage{graphicx}%
\usepackage{multirow}%
\usepackage{amsmath,amssymb,amsfonts}%
\usepackage{amsthm}%
\usepackage{mathrsfs}%
\usepackage[title]{appendix}%
\usepackage{xcolor}%
\usepackage{textcomp}%
\usepackage{manyfoot}%
\usepackage{booktabs}%
\usepackage{algorithm}%
\usepackage{algorithmicx}%
\usepackage{algpseudocode}%
\usepackage{listings}%

\newcommand{\coo}{CO\textsubscript{2}}
\newcommand{\hh}{H\textsubscript{2}}
\newcommand{\nn}{N\textsubscript{2}}
\newcommand{\oo}{O\textsubscript{2}}
\newcommand{\so}{$S{\rm_0}$}
\newcommand{\ei}{$E{\rm_i}$}
\newcommand{\en}{$E{\rm_n}$}
\newcommand{\ts}{$T{\rm_s}$}
\newcommand{\eo}{$E{\rm_0}$}
\newcommand{\thi}{$\theta{\rm_i}$}

\begin{document}

\title[Article Title]{
The curious case of \coo{} dissociation on Cu(110)
}

\author[]{Saurabh Kumar Singh}\email{sksaurabh@tifrh.res.in}

\author*[]{\fnm{Pranav R.} \sur{Shirhatti*}}\email{pranavrs@tifrh.res.in}

\affil[]{Tata Institute of Fundamental Research Hyderabad, 36/P Gopanpally, Hyderabad 500046, Telangana, India}

\abstract{
Dissociation of \coo{} on copper surfaces, a model system for understanding the elementary steps in catalytic conversion of \coo{} to methanol has been extensively studied in the past. 
It is thought to be reasonably well-understood from both experiments and theory.
In contrast, our findings reported here suggest a  different picture.
Using molecular beam surface scattering methods, we measure the initial dissociation probabilities (\so{}) of \coo{} on a flat, clean Cu(110) surface under ultra-high vacuum conditions.
The observed \so{} ranges from $3.9\times10^{-4}$ to $1.8\times10^{-2}$ at incidence energies of 0.64 eV to 1.59 eV with a lower limit to dissociation barrier estimated to be around 2.0 eV, much larger than that understood previously.
We discuss the possible reasons behind such large differences in our results and previous work.
These findings are anticipated to be extremely important for obtaining a correct understanding of elementary steps in \coo{} dissociation on Cu surfaces.}


\maketitle

\section*{Introduction}
The environmental impact of \coo{} production by the use of fossil fuels is understood to be a key player contributing to global climate crisis  \cite{IPCC_report_2023}.
While this problem is extremely complex and multi-faceted in nature, one of the proposed strategies for dealing with this is that of Carbon Dioxide Capture, Utilization, and Storage \cite{jiang_CO2_methanol_ChemRev_2020}.
In this context, the process of conversion of \coo{} to methanol (CH$_3$OH) has been of particular interest.
The methanol produced can be used directly as a fuel and also as a feedstock for the chemical industry, thereby enabling carbon recycling and reducing the damaging impact of increasing \coo{} emissions \cite{olah_methanol_2013, jiang_CO2_methanol_ChemRev_2020}.

\coo{} being a stable molecule from a thermodynamic and kinetic standpoint \cite{freund_CO2_rev_1996}, its chemical transformation is challenging and requires the selection of proper co-reactants and catalysts in order to achieve sufficient efficiency.
Copper/Zinc oxide/Alumina support-based catalysts, with \hh{} and CO as the co-reactants are commonly used in industrial processes for \coo{} conversion to methanol.
Based on previous work, \coo{} has been identified to be the main source of carbon in methanol formation, and \coo{} dissociation on the catalyst surface is understood to be a key step in the overall reaction scheme \cite{chinchen_carbon_source_methanol_1987, chinchen_CO2_Methanol_rev_1987}.
Understandably, the interaction of \coo{} with well-defined Cu single crystals has been used extensively as a model system to gain insights into the elementary steps involved in this catalytic process.
Among different low index planes of crystalline copper surfaces, Cu(110) has been of particular interest as the trend in the catalytic activity is observed to be following the order: Cu(110) $>$ Cu(100) $>$ Cu (111) \cite{campbell_methanol_1996}. 
The energy barrier for \coo{} dissociation on Cu(110) and Cu(100) single crystal surfaces have been reported to be 0.64 eV and 0.96 eV, respectively \cite{nakamura_CO2_Cu110_1989, chorkendorff_CO2_Cu100_1992}.
These measurements were performed using clean single crystalline copper surfaces, exposed to high pressure of \coo{}. 
The O-atom coverage and the initial sticking probability (\so{}) resulting from \coo{} dissociation, measured at different temperatures, were used to determine the dissociation barrier.
The calculated dissociation barriers for Cu(110) and Cu(100), assuming \coo{} to be interacting with idealized flat single crystal surfaces, using density functional theory (DFT) based computational methods agree very well with the above values \cite{morikawa_CO2_Cu_2014, yimin_APXPS_DFT_2020}.
Interestingly, in the case of Cu(111), where the catalytic activity is much lower and as a result direct experimental results are not available the situation is not as clear. 
The reported values of the dissociation barrier obtained using DFT-based methods show a rather large spread ranging from 1.69 eV \cite{gokhale_mechanism_2008}, 1.33 eV \cite{morikawa_CO2_Cu_2014} and 0.93 eV \cite{yimin_APXPS_DFT_2020}.
Nonetheless, the overall trend in the computed dissociation barriers \cite{Morikawa_comparison_2004} is consistent with the experimental observations  
and with the general understanding that more open surfaces, such as Cu(110) will have higher activity when compared to their closed-packed counterparts.
This is further confirmed by the fact that both experimental and computational studies on high index planes of Cu crystals, where the step densities are expected to be much higher, exhibit much lower dissociation barriers \cite{somorjai_Cu311_1992, morikawa_CO2_Cu_2014, kim_Cu977_2023}.

Given the above considerations, it is tempting to think that the model system of \coo{} interacting with well-defined Cu single crystals is well understood and can serve as a platform for building our understanding of realistic catalytic processes.
However, a closer look at the existing literature shows that a few essential questions of fundamental importance have largely remained unanswered. 
These are mainly concerned with the precise magnitude of the dissociation probabilities, its dependence on incidence energy, and the magnitude of the dissociation barrier on the terrace and step sites.
Having such information is of crucial importance in order to validate the prevailing microscopic picture underlying the \coo{} dissociation on Cu surfaces and to validate the estimates obtained from theoretical/computational approaches.

In light of the above considerations, we have carried out detailed measurements of the dissociation probabilities of \coo{} on Cu(110), using molecular beams under UHV conditions.
Absolute dissociation probabilities, measured as a function of the incidence energy are presented along with an estimate of the lower bound to the dissociation barrier.
Strikingly, our results show that the dissociation barrier is significantly higher, by at least about 3 times, when compared to the currently accepted value.
We present the likely hypotheses explaining these large deviations, along with a discussion of the broader implications of our results on the prevailing understanding of \coo{} dissociation on Cu surfaces in general.

\section*{Results and Discussion}\label{sec2}

\begin{figure*}[ht!]
\includegraphics[width=1\linewidth]{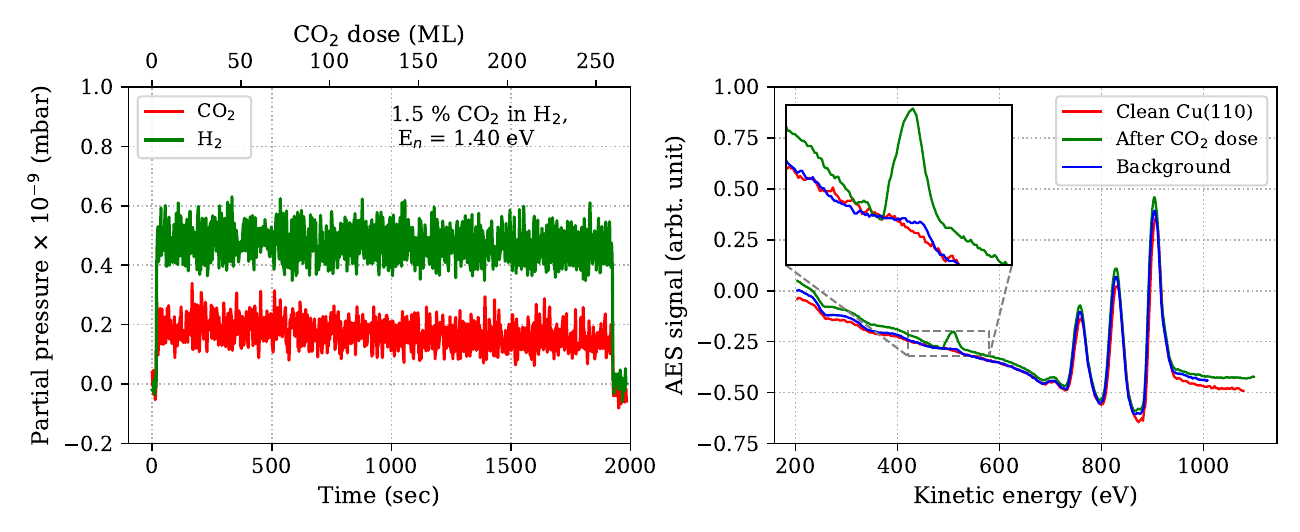}
\caption{(left) Partial pressure changes in the UHV chamber, monitored using a mass spectrometer when the incident molecular beam was turned on.
(right) Auger electron spectra of the Cu(110) surface measured after cleaning (red), after CO$_2$ dosing of 250 ML on an annealed surface (green). The background oxygen coverage build-up, measured at a 3 mm distance away from the position of \coo{} dosing is shown in blue. Peaks at 503 eV correspond to adsorbed oxygen, while peaks in the 700–920 eV region correspond to Cu. 
Inset shows a zoomed-in view of the oxygen peaks after dosing.
The maximum background oxygen build-up was estimated to be $<$ 5\% of saturation coverage in all our measurements.}
\label{fig1}
\end{figure*}

The present studies of the \so{} on a Cu(110) surface were performed using molecular beam-surface scattering (see methods).
Figure \ref{fig1} (left) shows an example of \coo{} and \hh{} partial pressure changes, observed in the UHV chamber housing the Cu(110) single crystal, upon turning on the molecular beam. 
In this example shown, a molecular beam of 1.5 \% \coo{} seeded in \hh{}, with an estimated incidence energy (\ei{}) of 1.40 eV was used. 
Auger electron spectra (AES), before and after exposing a clean Cu(110) surface to the molecular beam of 1.5 \% \coo{} in \hh{} (250 ML dose, surface temperature, \ts{} = 300 K) is depicted by the red and green curves, respectively (right panel).
A clear increase in the AES signal at 503 eV was observed (inset) after exposure to the \coo{} beam, indicating a build-up of O-atom coverage resulting from the dissociation of incident \coo{} on the surface.
Since \oo{} is also known to react readily with Cu(110) surface with a reported \so{} of 0.23 (\ei{} $<$ 50 meV) \cite{gruzalski_xps_1985, pudney_O2_Cu110_1990, nesbitt_O2_Cu110_1991} we also checked for any O-atom coverage build-up caused by background gas in the course of our measurements.  
This was estimated by measuring the AES signal at a nearby location (3 mm away from the dosing region) on the crystal, not exposed to the \coo{} beam (blue), measured at the end of the last dosing cycle (figure \ref{fig1}, right).
Throughout this study, we carefully monitored background oxygen build-up in each set of measurements, and its values were observed to remain below 5\% of saturation O-atom coverage (see SI-1).
All the O-atom coverage curves shown here subsequently have been corrected for this small background signal.
Further, by employing a pure beam of \coo{}, we observed no build-up of oxygen coverage, thereby ruling out any noticeable oxygen contamination in the incident molecular beam.
Additionally, a small carbon coverage (272 eV) (presumably due to background hydrocarbon adsorption), was also observed at long dosing times of the order of $2 \times 10^3$ seconds. 
We estimate the maximum carbon coverage in such cases to be less than $\sim$ 2\% of ML (see SI-2). 
Given its small value, we assume it to be not of much consequence for the measurements of \coo{} dissociation under consideration.

\coo{} dissociation on clean a Cu(110) surface  will result in CO and O formation. 
Given that the CO molecules are known to desorb from Cu(110) surface at temperatures $>$ 200 K \cite{burghaus_CO_Cu110_2001}, and that oxygen binds very strongly to the Cu(110) surface and the adsorbed layer remains intact even at much higher surface temperatures $>$ 770 K \cite{lapujoulade_heliumScat_OCu110_1982, somorjai_Cu311_1992}, we conclude that this O-atom coverage build-up results from dissociation of incident \coo{}.
Finally, by measuring the surface O-atom coverage as a function of the incident CO$_2$ dose (see methods section and SI-3 to SI-5) we estimated the initial sticking probability (\so{}) for \coo{} dissociation at different incidence energies.
\begin{figure*}[ht!]
\includegraphics[width=1\linewidth]{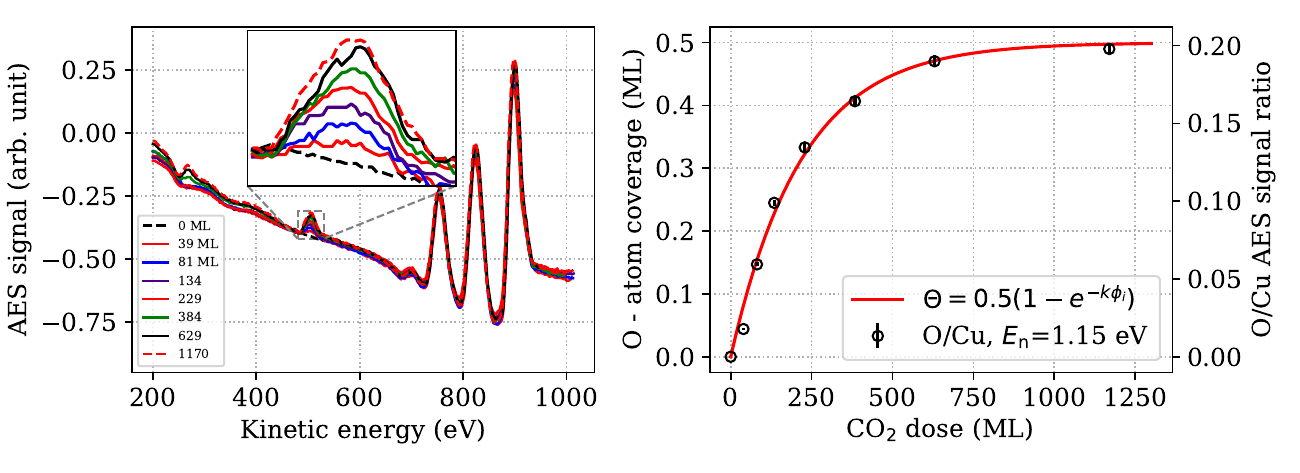}
\caption{(left) Auger electron spectra of the Cu(110) surface measured at different incident dose of \coo{}. (right) O-atom coverage build-up with increasing \coo{} dose. 
The coverage estimation was made using the AES peak ratio of O(503 eV) and Cu(776 eV).
The red curve is the best fit using a first-order kinetics model. 
}
\label{fig2}
\end{figure*}
Figure \ref{fig2} (left), shows the AES signal measured as a function of \coo{} dose (\ei{} = 1.15 eV, incidence angle, \thi{} = $0^{\circ}$), ranging from 0 ML (clean surface) to 1170 ML (saturation). 
A clear trend of increasing surface O-atom coverage with \coo{} dose can be seen. 
A quantitative analysis of this trend was obtained by plotting the ratio of oxygen to Cu peak-to-peak signal (background subtracted) as a function of incident \coo{} dose and is depicted in the figure \ref{fig2} (right). 
Notably, the ratio (O/Cu) reaches a value of 0.205 $\pm$ 0.005 at saturation coverage, which the is same as that obtained by dosing pure \oo{} (until saturation) on the same surface, measured independently.
Based on several previous studies using AES and low energy electron diffraction it has been well-established that saturation O-atom coverage corresponds to 0.5 ML, owing to a 2$\times$1  structure of the O-covered Cu(110) surface \cite{gruzalski_LEED_1984, gruzalski_xps_1985}.
This firmly establishes that the O-atom coverage observed under our measurement conditions remained unaffected due to CO + O recombination reaction and any unwanted reactions caused by the carrier gas (\hh{}) or the background gas.
Further, given that the saturation coverage corresponds to 0.5 ML, we convert the ratio of AES signals to surface O-atom coverage ($\Theta$), as shown in figure \ref{fig2} (right).
Here, the surface atom density of the Cu(110) was assumed to be $1.08 \times 10^{15}$ atom/cm$^2$  \cite{zhai_Cu110_density_2004}.
The surface O-atom coverage build-up as a function of the incident \coo{} dose was observed to be consistent with a simple first-order kinetics model.
This can be described by the equation $\Theta = 0.5(1-e^{-k \phi_{\rm i}})$, where $\phi_{\rm i}$ corresponds to the incident \coo{} dose (time-integrated incident flux) with the value of saturation coverage set to 0.5 ML. 
The slope of this function in the zero coverage limit ($ 0.5 \times k$) gives the initial dissociative sticking probability (\so{}) of CO$_2$ on Cu(110).

\begin{figure}[ht!]
\includegraphics[width=1\linewidth]{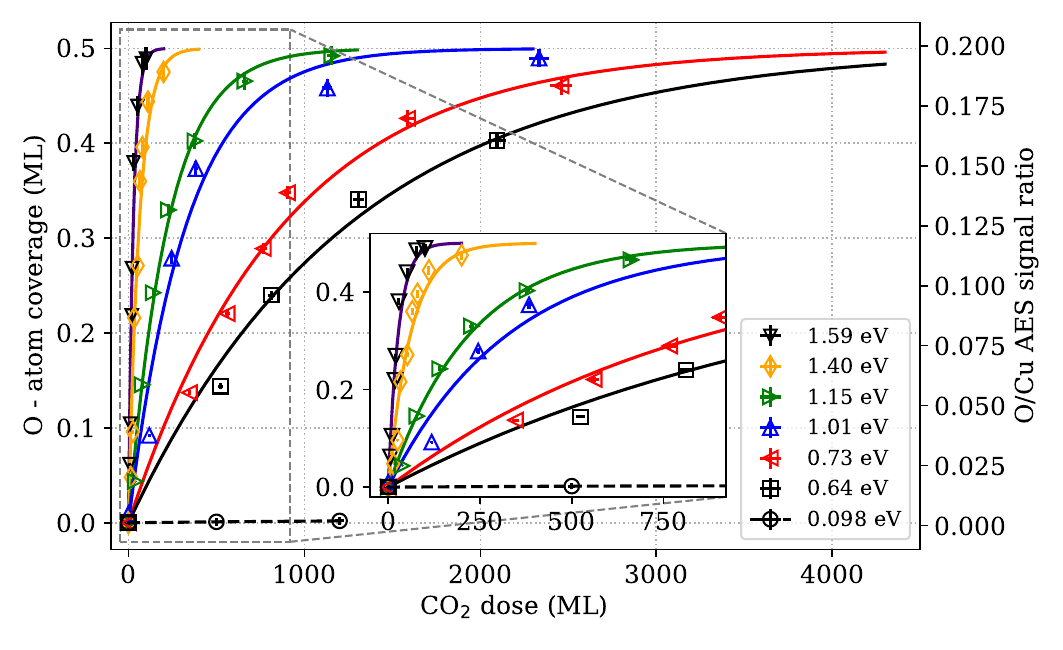}
\caption{A combined plot depicting O-atom coverage build-up (in ML) on Cu(110) surface as a function of incident \coo{} dose (in ML), measured for different translational energies. 
The inset shows a zoomed-in view of the initial kinetics.
At the lowest energy, we are unable to observe any O-atom coverage build-up.
Clearly, the \so{} increases with incident translational energy indicating a translationally activated dissociation.
} 
\label{fig3}
\end{figure}

\textbf{Incident translational energy dependence:} 
Dissociative chemisorption of CO$_2$ on Cu(110) was investigated over a range of translational energies spanning from 0.098 eV (100 \% \coo{}) to 1.59 eV (0.75 \% \coo{} in \hh{}).
Figure \ref{fig3} shows surface O-atom coverage (in ML) measured as a function of incident CO$_2$ dose (in ML) for seven different translational energies along with the best-fit curves.
These measurements were carried out with \ts{} = 300 K and \thi{} = $0^{\circ}$.  
As seen in figure \ref{fig3}, all the curves (except \ei{} = 0.098 eV) follow a similar pattern and approach the same saturation level of 0.5 ML O-atom coverage.
For the lowest incident energy of 0.098 eV, even upon dosing the surface with 1200 ML of \coo{}, the O-atom coverage remained indistinguishable from the background. 
Hence we conclude that the \so{} at 0.098 eV is below the detection sensitivity of our measurements, approximately $1.6 \times 10^{-5}$, limited by background oxygen coverage build-up in long dosing experiments.
Most importantly, with increasing \ei{} the initial slope increases, which is a clear signature of translationally activated dissociation.

\begin{figure}[ht!]
\includegraphics[width=1\linewidth]{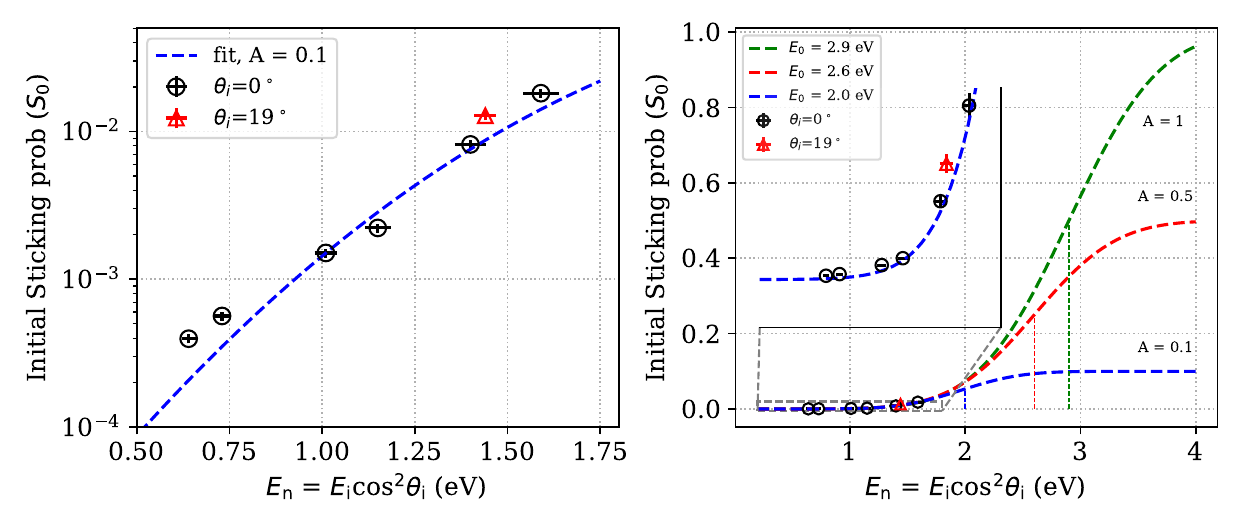}
\caption{
(left) A plot of \so{} obtained at different \en{} of \coo{}.
Black circles show the measurements at normal incidence and the red triangle corresponds to a measurement at $\theta_i$ = 19$^\circ$. 
(right) The same points (on a linear scale) are shown along with the best fits using S-shaped curves with different values of A of 0.1 (blue), 0.5 (red), and 1 (green).
Even for the lowest A (0.1), the \eo{} comes out to quite large and is 2.0 eV.
}
\label{fig4}
\end{figure}

The \so{} values derived from the initial slopes of the curves in figure \ref{fig3} are plotted against the translational energy associated with the normal component of incident momentum to the surface (\en{}) in figure \ref{fig4} (left).
Black circles depict \so{} values for measurements carried out at $\rm \theta_i$ = 0$^\circ$ and the red triangle refers to that obtained at $\rm \theta_i$ = 19$^\circ$ (\ei{} = 1.59 eV, \en{} = 1.42 eV). 
The blue dashed curve depicts an empirical fit function in the form of an S-shaped curve (discussed below). 
With increasing \en{} in the range of 0.64 to 1.59 eV, the \so{} increased from 3.9$\times$10$^{-4}$ to 1.8$\times$10$^{-2}$ (also see table \ref{tab_s0}). 
The measurement at $\rm \theta_i$ = 19$^\circ$ (red triangle) is consistent with the trend seen for the measurements performed at normal incidence, suggesting that only the normal component of the momentum (and the associated translational energy) is relevant for overcoming the dissociation barrier. 
This indicates that a simple one-dimensional barrier model can be used to understand this system.
In this case, the overall sticking probability can be expressed as:
\begin{equation}
S_0(E, T) = \sum_{v}F_{\rm B}(v, T) \cdot S_{0}(v)
\label{eq_overall_So}
\end{equation}
Where $F_{\rm B}(v, T)$ is the population in different vibrational states at a given vibrational temperature ($T$) of the incident beam, and $S_0(v)$ is the vibration state-specific initial dissociation probability.
Given that in our experiments, the nozzle is at room temperature (300 K), to a good approximation $F_{\rm B}(v = 0, T) = 1$, i.e. the population from the higher vibration states can be considered to be much smaller than the ground state.
The expression for $S_{0}(v)$ is given by a S-shaped curve (equation \ref{eq_scurve}). 
Here, the saturation value of \so{} is given by $A$, \eo{} corresponds to the dissociation barrier height (also the midpoint of the curve) and $W_0$ describes the distribution of the barrier heights.
\begin{equation}
S_{0}(v) = \frac{A}{2}\Biggl\{1+erf\Bigg[\frac{E_n-E_0}{W_0}\Bigg]\Biggr\}
\label{eq_scurve}
\end{equation} 
Given that, even at the highest incidence energy used in our measurements, the reaction probability is far from reaching its maximum value (it is still increasing) a precise estimation of the best-fit parameters is not possible at this stage.
Nonetheless, reasonably good estimates for the lower and upper limits of the barrier height can still be made (see figure \ref{fig4}, right).

If one considers the maximum value of \so{} to be similar to that observed here, the activation barrier comes out to be around 1.4 eV. 
However, given that the \so{} is still increasing even at the highest \en{} used, a more realistic estimate of the maximum \so{} would be around 0.1. 
The resulting \eo{} in this case is 2.0 eV (blue dashed curve, figure \ref{fig4}, right).
Further, assuming that the maximum \so{} to be 0.5 (red curve) and 1 (green curve), the estimated dissociation barriers come out to be 2.6 and 2.9 eV, respectively. 
In summary, even at incidence energies $>$ 1.5 eV, the \so{} values are low and of the order of $10^{-2}$, and the dissociation barrier is estimated to be at least of the order of 2 eV. 
It is worth pointing out here that according to previous work \cite{burghaus_CO2_Cu110_2006}, at \en{} = 1.3 eV, using a heated nozzle at 750 K, no dissociative chemisorption was observed.
Even more striking is the fact that the dissociation barrier estimated from our observations is much larger when compared to 0.64 eV reported previously \cite{nakamura_CO2_Cu110_1989, yimin_APXPS_DFT_2020}. 
In the following discussion, we present a few hypotheses based on which these large differences can be possibly rationalized.

\begin{table}[h]
\caption{Gas mixture composition, \en{} and the observed \so{} values as shown in figure \ref{fig4}. All measurements were performed at normal incidence except for one shown in the second row.
The random uncertainties in the \so{} values were evaluated to be about 14\% (see methods and table \ref{tab_error} for a discussion on uncertainties).
}
\label{tab_S0}
\vspace{.5cm}

\centering 
\begin{tabular}{ p{4cm} p{2cm} p{2cm}} 
\hline\hline 
Gas mixture composition & \en{} (eV)  & Initial sticking probability \\ [0.5ex]
\hline \\

0.75 \% \coo{} in \hh & 1.59 & $1.8 \times 10^{-2}$  \\[1ex]

0.75\% \coo{} in \hh{} ($\theta_i = 19^{\circ}$) & 1.43 & 
$1.3 \times 10^{-2}$  \\[1ex]

1.5\% \coo{} in \hh{} & 1.40 & $8.1 \times 10^{-3}$  \\[1ex] 

2.9\% \coo{} in \hh{} & 1.15 & $2.2 \times 10^{-3}$  \\[1ex] 

4\% \coo{} in \hh{} & 1.01 & $1.5 \times 10^{-3}$  \\[1ex] 

7.5\% \coo{} in \hh{} & 0.73 & $5.6 \times 10^{-4}$  \\[1ex] 

9.2\% \coo{} in \hh{} & 0.64 & $3.9 \times 10^{-4}$  \\[1ex]

100\% \coo{} & 0.098 &  $< 1.6 \times 10^{-5}$  \\[1ex] 
\hline 

\end{tabular}
\label{tab_s0}
\end{table}

\textbf{Understanding the activated dissociation of \coo{}:}
First, we provide a detailed comparison of our results with those reported previously by Funk and co-workers  \cite{burghaus_CO2_Cu110_2006}. 
As both these studies were carried out using molecular beams under UHV conditions, a systematic comparison is relatively easier to make.  
The main objective of this previously reported study was to understand the physisorption dynamics of \coo{} on the Cu(110) surface. 
As a consequence, a cold surface below the desorption temperature of \coo{} (90 K) was used in their work, as opposed to the surface being at 300 K in our work.
Under these conditions, they report an \so{} for the non-dissociative physisorbed \coo{} to be 0.05 (\en = 1.3 eV), which is about a factor of 10 higher than \so{} for dissociative chemisorption at the same incidence energy (see figure \ref{fig4}, left).
Since physisorbed \coo{} will stay on the surface in measurements made below its desorption temperature, and that its \so{} (non-dissociative) is much higher, a very small fraction of surface sites are expected to be available for dissociative chemisorption of the incoming \coo{}. 
Under such conditions, it is very likely that the dissociative chemisorbed signal will be very small and remain below the detection threshold in their measurements, reported to be approximately 0.03 ML.

\begin{figure}[ht!]
\includegraphics[width=.9\linewidth]{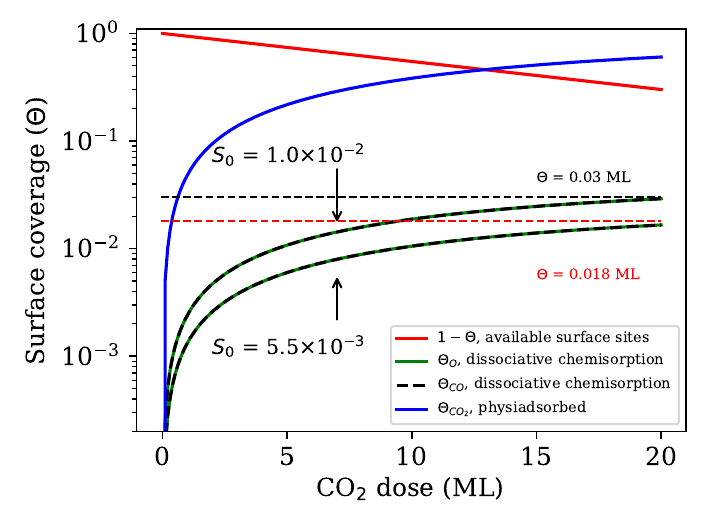}
\caption{Surface coverage for O-atom, CO, and \coo{} calculated using the kinetic model for a low-temperature surface. 
The \so{} values for \coo{} physisorption and dissociative sticking are used are $5\times10^{-2}$ and $5.5\times10^{-3}$, respectively.
The red dashed horizontal line shows maximum surface coverage arising from \coo{} dissociation assuming a \so{} of $5.5\times 10^{-3}$ (present work).
The black dashed horizontal line shows the detection sensitivity of the surface coverage due to \coo{} dissociation from previous work \cite{burghaus_CO2_Cu110_2006}. 
As can be seen from the green and black dashed curves, even if the \so{} is as high as $1.0\times10^{-2}$ (say due to additional contribution of vibrationally excited \coo{} in the incident beam), the resulting surface coverage will remain close to the detection threshold of 0.03 ML reported previously \cite{burghaus_CO2_Cu110_2006}.  
}
\label{fig5}
\end{figure}

A simple first-order kinetics-based model was used to estimate the expected surface coverage based on the \so{} values available for non-dissociative physisorption (see SI-6) and dissociative chemisorption of \coo{} (present work).
Since the non-dissociative physisorption decreases with \en{} and dissociative chemisorption increases with \en{}, it is useful to make this comparison at the highest energy (1.3 eV) used in their work.
With \so{} for the dissociative chemisorption as $5.5\times10^{-3}$ and that for non-dissociative physisorption as $5\times10^{-2}$ the kinetic model predicts the maximum surface coverage due to due to dissociative chemisorption to be less than 0.018 ML (see figure \ref{fig5}).
This is well below the reported detection limit of 0.03 ML, possibly explaining the absence of any signature of dissociative chemisorption in their measurements. 

It is worth pointing out that a heated nozzle (750 K, for \ei{} = 1.3 eV) was used in the previous experiments \cite{burghaus_CO2_Cu110_2006} and an additional contribution to the dissociative sticking channel due to vibrationally excited molecules in the incident beam is possible.
Based on the reported detection sensitivity ($\Theta= 0.03$ ML) and the fact that they were unable to see any dissociation also allows us to estimate an upper bound to the vibrational enhancement of dissociative chemisorption of \coo{} on Cu(110).
Again using the same kinetics model, we estimate this upper limit to the \so{}, using a hot nozzle at 750 K, to be $1.0\times10^{-2}$ at 1.3 eV.
This is approximately a factor of 1.8 times higher than that obtained with a room-temperature nozzle in our studies.
This is of significance as several studies have concluded that the transition state for \coo{} dissociation is via a bent configuration \cite{morikawa_CO2_Cu_2014,yimin_APXPS_DFT_2020} and in a molecular beam using a hot nozzle, a significant amount of bending mode excited \coo{} will be present in the incident beam.
At a nozzle temperature of 750 K and assuming negligible vibrational relaxation in the supersonic jet expansion, the population of bending mode excited \coo{} will be about 21\%. 
Based on this we estimate the maximum \so{} for purely bending excited \coo{} to be approximately $2\times10^{-2}$.
Corresponding vibrational efficacy \cite{beck_tutorial_2016}, the ratio of translational energy to vibrational energy needed to reach the same \so{},
assuming that the bending mode excitation is solely responsible for enhanced reactivity, can be as large as 5 (upper limit) for this system.
Similar observations in the case of \coo{}/Ni(100) have been reported by \cite{madix_CO2_Ni100_1986}, where for given \en{} with increasing nozzle temperature a clear enhancement in \so{} was observed.
This was attributed to the significantly higher dissociation probability of incident molecules with vibration excitation, in particular with the bending mode excitation.
Recent theoretical studies on the same system \cite{guo_CO2_Ni100_2016, jackson_CO2_Ni100_2017} are also in agreement with the conclusions.

We now turn our attention towards a comparison of our results with those reported previously using high-pressure measurements \cite{nakamura_CO2_Cu110_1989}.
Based on exposing the Cu(110) surface with \coo{} gas at 65 and 650 Torr pressure, under well-controlled conditions such that any effect of contamination giving rise to spurious oxygen coverage is minimized, they observe O-atom coverage build-up on the surface using AES and conclude that this results from \coo{} dissociation.
The O-atom coverage was reported to increase with duration of \coo{} exposure (dose), and also increase with increasing surface temperature.
They estimate the dissociation probability to be  $1\times10^{-11}$ to $1\times10^{-9}$ with the surface temperature ranging from 430 K to 612 K, respectively.
Based on the temperature-dependent dissociation rates, the activation barrier was estimated to be 0.64 eV in the low coverage limit.
More recent studies on the same system \cite{yimin_APXPS_DFT_2020}, also compare favourably.
Using the near-ambient X-ray photoelectron spectroscopy (NAXPS) method at 300 K and exposure to 1 mbar \coo{} gas, a reaction probability of $4.4\times10^{-11}$ per collision on the surface was reported.
One of the peaks observed in NAXPS is attributed to be the signature of the molecularly chemisorbed \coo{}, which is understood to be anionic in nature with a bent structure.
Additionally, they also find that using DFT-based computational methods the activation barrier comes out to be 0.64 eV, consistent with those reported earlier. 
At the same time, it is also quite clear that these results are inconsistent with those obtained using molecular beams previously \cite{burghaus_CO2_Cu110_2006} and the present work.
If indeed, the dissociation barrier was as low as 0.64 eV, much higher dissociation probabilities would have been observed in these measurements, especially at incidence energies as high as twice the dissociation barrier.
One possible way to understand this would be that the reaction follows completely different pathways under high-pressure conditions as opposed to that using molecular beams. 
In the former case, given that the translational energy is low, most molecules will undergo trapping on the surface and consequently, the reaction will proceed via a precursor-mediated pathway. 
The trapped molecules on a hot Cu(110) surface will thermalize and acquire the chemisorbed state (bent structure) and subsequently follow a low activation barrier pathway for dissociation.
On the other hand, such a low-energy pathway might be inaccessible to the incident \coo{} molecules from the gas phase, as in the case of molecular beams.
Here, the reaction will follow a direct dissociation pathway which is highly translationally activated as observed in our measurements.
A deeper understanding of this issue can be obtained by means of trajectory calculations, as reported previously in the case of \coo{} on Ni(100) and W(110) surfaces \cite{guo_CO2_Ni100_2016, guo_CO2_W110_2022}. 

Another possibility for explaining this large difference would be that at low vs. high-pressure conditions, fundamentally different surface structures are present leading to different reactivity. 
Such differences are commonly termed as \textit{pressure gap} and can lead to systematic differences among the high-pressure and low-pressure studies.
As an example, Eren and coworkers \cite{somorjai_CO2_Cu_reconstruction_2016, somorjai_reconstruction_CO_Cu111_2016}  have studied the interaction of Cu(100) and Cu(111) surfaces with \coo{} and CO under high-pressure conditions.
In the case of \coo{} at pressures beyond 20 Torr, they conclude that the surface breaks up into nano-sized clusters producing highly reactive kink and step sites.
They also report that the saturation O-atom coverage under these conditions is higher than that observed at lower pressures, where no such restructuring occurs. 
It should be noted that the saturation surface O-atom coverage in our measurements and in that reported under high-pressure conditions are both the same (0.5 ML). 
This suggests that in both these cases (high and low-pressure studies) the surface structure is likely to be similar and this alone can not be the reason for such large deviations observed in the dissociation barriers.

Finally, large differences in the activation barriers can arise from different reaction sites, i.e. steps vs. terraces in high vs. low pressure studies, respectively. 
This situation is reminiscent of that reported for \nn{} dissociation on Ru(0001) \cite{chorkendorff_N2_Ru_PRL_1999, chorkendorff_N2_Ru_2000}.
In these studies, the dissociation barrier determined using high-pressure reactions was reported to be 0.4 eV, whereas that obtained using molecular beam studies was observed greater than 1 eV \cite{luntz_N2_Ru_2001}.
It was also reported that upon blocking the step sites by adsorbing a small fraction of Au atoms on the surface (1-2\% ML) a remarkable drop in the reactivity by a factor of $10^9$ was observed.
At the same time, the corresponding decrease observed in the molecular beam measurements was reported to be much smaller, only by a factor of two.
Based on these observations and the fact that the step density on their surface was estimated to be $<$ 1\%, it was concluded that under high-pressure conditions, \nn{} mainly dissociates on the step sites, which are much more active than terrace sites. 
This is followed by the diffusion of N-atoms to the terrace sites, thereby allowing the reaction to proceed further.
On the other hand in molecular beam experiments, the \nn{} mainly dissociates on the terrace sites which are available in a much larger fraction. 

Given the above considerations, it is very likely that a similar situation is prevailing in the case of \coo{} on Cu(110).
Even a small fraction of the steps ($\sim$1\%) having much higher reactivity could lead to a much lower activation barrier being observed in the high-pressure experiments as compared to that using molecular beams.
This is also consistent with previous studies where using carefully prepared surfaces with higher step densities, more facile dissociation was observed \cite{somorjai_Cu311_1992, kim_Cu977_2023}.
In a more closely related system of \coo{} dissociation on Cu(100) surface, recent studies \cite{hagman_steps_2018} using APXPS and DFT-based computational methods it was inferred that the steps play a very important role in the reaction under high-pressure conditions.
In a recent DFT-based study carried out by Jin and coworkers  \cite{guo_CO2_diss_step_vs_flat_2022} \coo{} interaction with a very large range of metal surfaces, both flat and stepped, was studied. 
The general trends observed here too show the much higher activity of the step sites for \coo{} dissociation.
We have made a preliminary attempt to understand this by measuring the \so{} on a clean sputtered surface and comparing it with an annealed surface (see SI-7).
However, the changes observed are too small to be conclusive at the moment, and further systematic studies will be needed.

It would also be very interesting to look into \coo{} dissociation on Cu(100) and Cu(111) surfaces, using molecular beam techniques such as those presented here, so that the dissociation barrier corresponding to terrace sites can be measured unambiguously.
Finally, we would like to point out that the dissociation barriers based on DFT studies are reported to be 1.69 eV - 0.97 eV, 0.93 eV, and 0.64 eV on Cu(111), Cu(100), and Cu(110) respectively.
While the trend observed is consistent with that known in experiments, our present findings strongly suggest that these are likely to be severely underestimated.

\section*{Conclusion}\label{sec13}
In summary, using molecular beam methods we report that the \coo{} dissociation barrier on Cu(110) terrace sites is of the order of 2 eV, much higher than that known previously. 
Among different possible reasons, these observations suggest that there could be a large impact of step sites in driving dissociation under high-pressure conditions, resulting in a substantially low activation barrier when compared to terrace sites, as observed using molecular beams. 
This insight prompts into the direction where a critical reevaluation of the microscopic picture associated \coo{} dissociation on Cu surfaces is necessary.
Specifically, one needs to carefully reexamine the barrier heights and the corresponding dissociation probabilities on step versus terrace sites on low-index copper surfaces.
Our study also suggests that it will be very interesting to look into the reactivity of vibrationally excited \coo{} in order to understand the mode specificity of this reaction.
The estimates of the vibrational promotion in \coo{} dissociation on Cu(110) provided here will need to be tested using hot nozzle/infra-red excitation methods. 
In case significant vibrational promotion is found to be absent, it would again point towards the possible role of steps in \coo{} dissociation on Cu(110) under high-pressure conditions, leading to a much lower dissociation barrier.
Additionally, these results also suggest that it is crucial to carefully evaluate the surface diffusion barriers, especially the step to terrace migration, in order to fully understand the microscopic details involved in case of high-pressure conditions.
In the near future, we will be looking into some of these aspects in order to obtain a deeper understanding of the surface chemistry of \coo{} on Cu surfaces.

\section*{Methods}\label{sec11}

Our experiments were conducted using a recently designed molecular beam-surface scattering apparatus. It consists of a source chamber, and two differential pumping stages (Diff-1 and Diff-2) followed by a UHV chamber where the Cu(110) single crystal is placed. 
The source, Diff-1 Diff-2, and UHV chamber were pumped using a turbomolecular pump with nominal pumping speeds of 1200 l/s (HiPace 1200, Pfeiffer Vacuum), 400 l/s (HiPace 400, Pfeiffer Vacuum), 80 l/s (HiPace 80, Pfeiffer Vacuum) and 700 l/s (HiPace 700H, Pfeiffer Vacuum), respectively. The turbomolecular pump for the source and Diff-1 were backed by a 35 m$^3$/hour two-stage rotary vane pump (Duo 35, Pfeiffer Vacuum), and the Diff-2 and UHV stages were backed by a dry roots pump (ACP 15, Pfeiffer Vacuum).
A pulsed solenoid valve with an opening diameter of 1 mm (Parker 009-1643-900, driver IOATA ONE 060-0001-900) placed in the source chamber was used as a molecular beam source.
The supersonically expanded gas was made to pass through a 1.5 mm opening diameter skimmer (Beam Dynamics) and two subsequent apertures (2 mm diameter) placed downstream at the entrance of Diff-2 and UHV chambers.
The overall source-to-sample distance was approximately 340 mm. 
The beam diameter at the target surface was measured to be 2.9 mm (see SI).
The ultimate base pressure of the UHV chamber, monitored using a nude ion gauge (IMR 430, Pfeiffer vacuum) ranged from (6 to 8) $\times 10^{-10}$ mbar. 
This ion gauge was independently calibrated by \so{} measurements of \oo{} on a clean Cu(110) surface, measured using the molecular beam reflection method \cite{king_wells_S0_1972} (see SI-3). 
An Ar ion source (IS40, Prevac) was used to clean the copper surface by sputtering and an Auger electron spectrometer (SMG600, OCI Vacuum Microengineering) was used to analyze the surface chemical composition. 
Additionally, a mass spectrometer (SRS RGA 200), calibrated with the ion gauge as a reference, was used to measure the residual gas composition as well as for the estimation of the incident beam flux.

With the \coo{} molecular beam on, the pressure in the source, Diff-1, Diff-2, and UHV in the chamber typically was to ($0.5-2)\times10^{-4}$, $(1-3)\times10^{-6}$ mbar, $(5-8)\times10^{-7}$ mbar and $(4-7)\times10^{-9}$ mbar, respectively. 
The purity levels of the gases used in our measurements were specified to be $>$99.999\% for \hh{} and $>$ 99.99\% for \coo{} and were used without any further treatment.
A Cu(110) single crystal 99.9999 \% pure, 10 mm diameter, and 2 mm thickness), cut to an accuracy better than 0.1 and polished to have a roughness lower than 10 nm (MaTeck Material Technologie and Kristalle GmbH), was used as a target sample. 
It was mounted on a four-axis differentially pumped manipulator using a pair of 0.25 mm diameter tungsten wires that enabled the sample heating.
The sample manipulator is equipped with electrical and thermocouple feed-throughs for heating the sample and monitoring its temperature using a K-type thermocouple.
The \coo{} flux on the target surface ranged from 0.05 - 0.8 ML/sec, where 1 ML corresponds to $1.08\times10^{15}$ atoms cm$^{-2}$. 
Throughout all measurements, the backing pressure was maintained at a constant value of 5 bar, and the nozzle pulsing rate was typically 10 Hz. The nozzle opening time was varied within the range of 300 to 400 $\mu$s.

The Cu(110) surface was cleaned according to well-established procedures reported previously \cite{musket_surface_prep_1982, geetika_compact_He_2023}. 
After a bakeout of the UHV chamber, the main contamination at the copper surface was found to be carbon.
This carbon was removed by heating the Cu(110) surface (700 K) in an oxygen environment (at $2 \times 10^{-8}$ mbar). 
Subsequently, the remaining oxygen contamination was removed (as seen in AES) by prolonged Ar ion sputtering.
Thereafter, for day-to-day operation, the sample surface was subjected to Ar ion sputtering for a duration of 30 min (0.6 $\mu A$ ion current) at 3 keV ion energy. 
Under these conditions, impurity levels (mainly carbon and oxygen) were found to be below the detection threshold ($<$ 0.1\% ML and 2\% ML, respectively) of AES.
Subsequently, the surface was annealed at 800 K for 20–30 min and allowed to cool down to 300–310 K before conducting the measurements. 
With a base pressure of $8 \times 10^{-10}$ mbar, we observed that the impurity levels of carbon and oxygen (as measured by AES) remained $<$ 3\% of a ML for a duration four hours.
We also measured background build-up in each measurement at the end of the experiments (see SI-1) and these were found to be negligibly small in the timescale of our measurements.

Molecular beams with different translation energies were prepared by using different fractions of \coo{} seeded in \hh{}.
The incidence translational energy \coo{} in these gas mixtures was estimated using the following relation:
\begin{equation}
E_i = \frac{X_{CO_2}C_{P_{CO_2}} + X_{H_2}C_{P_{H_2}} }{X_{CO_2}{M_{CO_2}} +X_{H_2}{M_{H_2}}}M_{CO_2}(T_{N} - T_{R})
\label{eq: trans energy}
\end{equation} 
Here, X$_{CO_2}$ and X$_{H_2}$ represents the mole fraction of \coo{} and \hh{}, respectively. C$_{P_{CO_2}}$ and C$_{P_{H_2}}$ indicate the heat capacities of \coo{} and \hh{}, respectively. 
T$_{N}$ corresponds to the nozzle temperature, while T$_{R}$ represents the rotational temperature of the molecular beam.
By utilizing the parameters of T$_{N}$ = 300 K and assuming T$_{R}$ = 10 K (typical for molecular beams), we calculated the translational energy, E$_i$, of the CO$_2$ beam in a given mixture (see \ref{tab_s0}).
Based on previous work, we estimate that these calculated beam energies will be accurate within 10 \% of that reported here. 

The initial dissociative sticking probability of \coo{}  on Cu(110) was measured in the energy range of 0.098 eV to 1.59 eV. To determine the coverage of adsorbed oxygen resulting from dissociation, an Auger electron spectrometer was used with a 1 $\mu A$ surface current and a 2.5 keV beam energy.
Under these conditions, we could not observe any change in O-atom coverage caused by electron-stimulated desorption.
The coverage was determined by measuring the ratio of peak-to-peak heights at electron energies of 503 eV (O) and 776 eV (Cu) using the Auger electron spectrometer.
The observed ratio of 0.205 $\pm$ 0.005 indicated a saturated O-atom coverage of 0.5 ML on the Cu(110) surface. 
To estimate the incident \coo{} flux, a correlation between \coo{} and \hh{} was established by utilizing measurements from a calibrated ion gauge and a mass spectrometer. 
The gas-dependent sensitivity factors for the ionization gauge, 1.42 for \coo{} and 0.46 for \hh{}, as well as calibration factors for the ion gauge and mass spectrometer, were used for estimating the incident \coo{} flux (see SI-5). 
Additionally, we have excluded any significant contribution from potential contaminants present in the incident beam, such as CO or \hh{}O, which could possibly arise from the reverse water gas shift reaction (see SI-8).

\textbf{Uncertainty estimates:} 
The random errors in the \so{} estimation arose from the uncertainties in surface O-atom coverage and the incident beam flux estimation.
The contributing factors are listed in table \ref{tab_error}.
Since the pumping speeds of \coo{} and oxygen are not directly available from the manufacturer's datasheet, we have assumed them to be equal to gases with similar mass such as Argon (665 l/s) and nitrogen (685 l/s), respectively. 
These values are as per the manufacturer's specification and uncertainty arising due to deviations here will contribute mainly to the systematic errors, leaving the overall trends reported here unchanged.
\begin{table}[h]
\caption{A breakdown of the different contributing factors to the random uncertainties in \so{} estimation. 
$\delta f/f$ here represents the 1$\sigma$, relative errors given as a percentage}
\label{tab_error}
\vspace{.5cm}

\centering 
\begin{tabular}{p{5cm} p{1.5cm} p{6cm}}
\hline\hline 
Source of error &  $\delta f/f$ (\%) & Remarks  \\ [0.5ex]
\hline \\

Uncertainty in determining absolute pressure (ion gauge calibration) & 10\% & From \so{} measurements of \oo{} on Cu(110) (SI-3) \\[1ex]

Uncertainty in calibrating the mass spectrometer with ion gauge & 5 \% & From the uncertainty in the fit parameters describing the correlation among mass spectrometer and ion gauge signals (SI-5) \\[1ex] 

Uncertainty in beam shape estimation & 5\% &  Approximate estimate based on the measurements shown in SI-4 and assuming the shape to be same for all beams \\[1ex]

Uncertainty in AES signal determination & 5 \% & From statistics of repeated AES measurements \\[1ex]

Uncertainty in repeatability of sample positioning while dosing & 5 \% & Approximate estimate from beam shape estimation and assuming $\pm$0.25 mm sample positioning error \\[1ex]

Overall random uncertainty & 14\% & Assuming independent errors \\[1ex]

\hline 
\end{tabular}
\label{table1}
\end{table}

\bmhead{Supplementary information} 

\begin{itemize}
    \item SI-1: Estimates for background oxygen coverage build-up
    \item SI-2: Estimates for background carbon coverage build-up
    \item SI-3: Ion gauge calibration
    \item SI-4: Molecular beam profile at the Cu(110) surface
    \item SI-5: Estimating the flux of the incident beam
    \item SI-6: Kinetic model for estimating surface coverage caused by non-dissociative physisorption and dissociative chemisorption
    \item SI-7: Influence of defects on \so{}
    \item SI-8: Discussion on possible contamination in our incident beams
\end{itemize}

\bmhead{Acknowledgments}
We acknowledge the support of intramural funds at TIFR-Hyderabad provided by the Department of Atomic Energy, Government of India, under Project Identification No. RTI 4007 and Scientific and Engineering Research Board, Department of Science and Technology, India (Grant number. CRG/2022/002943). 
We thank Avinash Kumar for his help in setting up the molecular beam-surface scattering apparatus.

\bmhead{Data Availability}
All relevant data related to the current study are available from the corresponding author upon reasonable request.

\bmhead{Author contributions}
SKS and PRS conceived and designed the study. 
SKS performed the measurements and analyzed the results with inputs from PRS. 
SKS and PRS discussed the results and prepared the manuscript.

\bmhead{Conflict of interest}
The authors declare no conflict of interest.

\bibliography{bibliography}%

\end{document}


\title[Article Title]{The curious case of \coo{} dissociation on Cu(110): Supplementary information }

\author[]{Saurabh Kumar Singh}\email{sksaurabh@tifrh.res.in}

\author*[]{\fnm{Pranav R.} \sur{Shirhatti*}}\email{pranavrs@tifrh.res.in}

\affil[]{Tata Institute of Fundamental Research Hyderabad, 36/P Gopanpally, Hyderabad 500046, Telangana, India}

\maketitle

\tableofcontents

\newpage
\section{SI-1: Estimates for background oxygen coverage build-up}
\begin{figure}[H]
\includegraphics[width=1\linewidth]{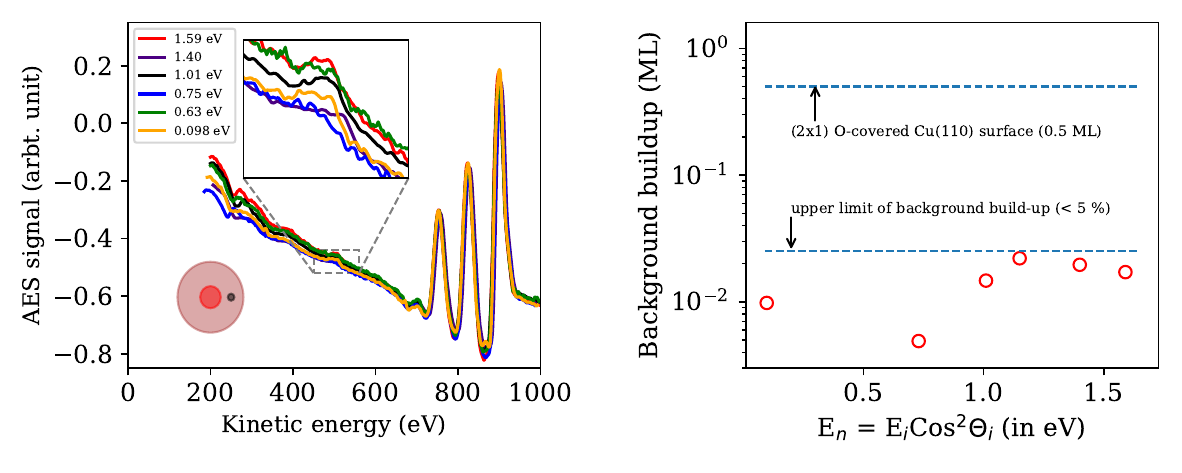}
\label{Fig_01}
\centering
\caption{(left) Auger electron spectra (AES) measured to estimate background O-atom coverage corresponding to each measurement at different incidence energies. 
A magnified view of the O-atom coverage buildup is shown in the inset. 
A schematic diagram indicating the different measurement positions is also shown (pink circle).
The molecular beam was incident on the center of the target surface (red spot) for the \so{} measurements.
All the background measurements were conducted at a 3 mm distance away (black spot) from the center of the target surface at the end of the final dosing cycle.
(right) The magnitude of the background O-atom coverage (in ML) for measurements carried out at different incidence energies.
The maximum background oxygen coverage observed was about 0.025 ML, corresponding to 5\% of the saturation coverage.}
\end{figure}

To assess the contribution of background O-atom coverage in our measurements, we use auger electron spectra of the Cu(110) surface, measured after the final dosing cycle in each measurement (see figure \ref{Fig_01}, left). 
This was measured at a location 3 mm away (black spot) from the point of impact of \coo{} molecular beam for \so{} measurements (red), as depicted in Figure 1 (left). 
Our analysis indicated that the background O-atom coverage amounted to less than 5\% of the saturation coverage (see figure \ref{Fig_01}, right) and is expected to have a negligible effect on the \so{} values reported in our work.

\newpage
\section{SI-2: Estimates for background carbon coverage build-up}

We have quantified the maximum background carbon coverage build-up based on the AES sensitivity factors and also by comparing with an unclean surface with carbon as the major contaminant. 
\cite{auger_database_1978,godowski1998augerdatabase}
The maximum background carbon coverage was observed to be $<$ 1.2\% of ML and is expected to have a negligible effect on the \so{} measurements reported in our work.
An example, the estimation of carbon coverage measurement is shown below in figure \ref{Fig_02}.

\begin{figure}[H]
\includegraphics[width=1\linewidth]{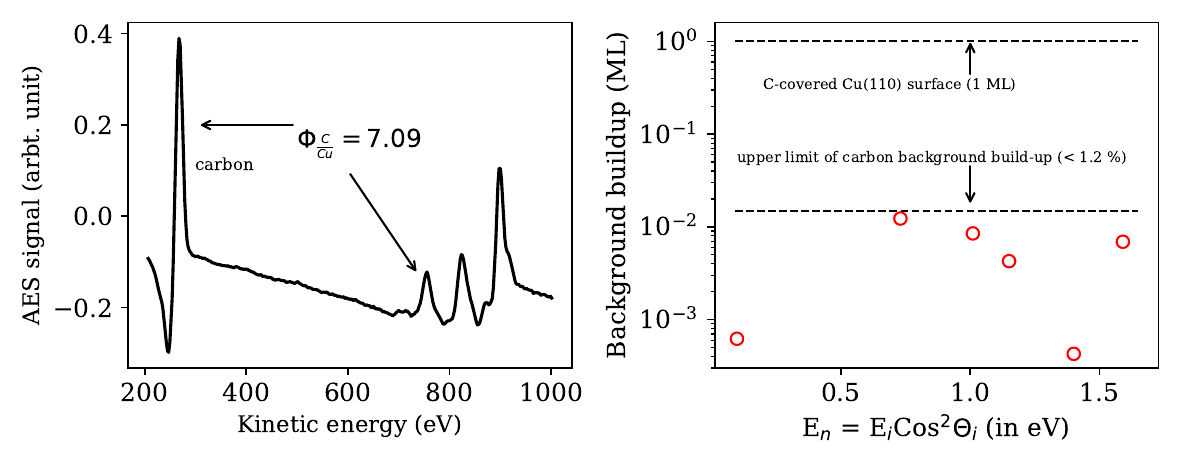}
\centering
\caption{(left) AES obtained from an unclean Cu(110) surface with carbon as the major contaminant.
AES signal ratio (C-272 eV)/(Cu-776 eV)  (marked with an arrow) ratio of 7.09 was observed.
(right) Estimates of the background carbon coverage in the reported \so{} measurements, assuming 7.09 to be the carbon saturation coverage.
In reality, the saturation carbon coverage is slightly higher (as per AES sensitivity factors), consequently, these values represent an upper limit to the background carbon coverage.
}
\label{Fig_02}
\end{figure}

\section{SI-3: Ion gauge calibration}

To accurately quantify the absolute reaction probability, estimation of the incident \coo{} dose is essential. 
In our study, we use an ion gauge and a quadrupole mass spectrometer for precise dose calculations. 
Calibration of the ion gauge was carried out by measuring the \so{} and O-atom coverage resulting from \oo{} dissociation on a clean Cu(110) surface, using beam reflection method \cite{king_wells_S0_1972}.
A beam of pure \oo{}, with an estimated incidence energy of 0.085 eV was used for this purpose.

It is well established that the \so{} and O-atom saturation coverage on a clean Cu(110) are 0.35 ± 0.03 (at 0.085 eV) and 0.5 monolayers (ML), respectively \cite{nesbitt_O2_Cu110_1991,gruzalski_LEED_1984}.
These values were used for the calibration of the ion gauge as follows:
The fraction of pressure reduced immediately when a pure \oo{} beam is incident on a clean Cu(110) surface gives the \so{}.
The overall reduction of the time-integrated oxygen pressure (measured using the ion gauge), until saturation coverage is reached and no further sticking is possible, corresponds to the number of molecules equivalent to the saturation coverage of 0.5 ML.  
This allowed us to calibrate the ion gauge.

\begin{figure}[H]
\includegraphics[width=1\linewidth]{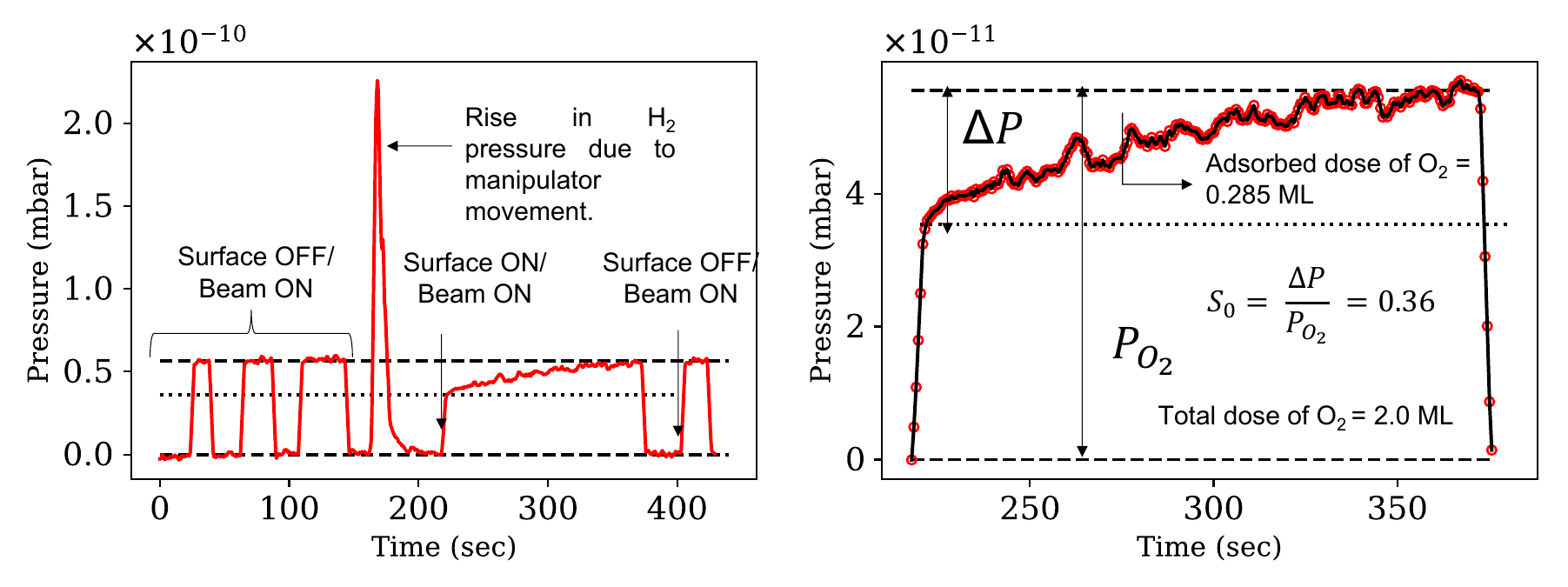}
\caption{(left) Temporal evolution of oxygen pressure is depicted measured using the ion gauge, with the molecular beam on and off.
The initial three cycles illustrate the repeatability of the pressure changes. 
Here, the surface was not in the line of sight of the molecular beam. 
The fourth measurement cycle was conducted while exposing the Cu(110) surface to the beam. 
The fifth cycle was measured to check for any systematic drifts.
The rapid increase in the pressure at 170 sec is caused by a jump in the \hh{} pressure resulting from by the movement of the sample manipulator and does not affect our measurements.
(right) A zoomed view of the fourth cycle shows the decrease in \oo{} pressure caused by sticking to Cu(110) surface.
The fractional decrease in \oo{} signal at initial times gives its \so{} and the overall decrease corresponds to the total number of molecules adsorbed on the surface.}
\label{FIG_03}
\end{figure}

Figure \ref{FIG_03} shows the result of one such measurement.
First, we checked the repeatability of the pressure changes in our system for three on-off cycles of the pure \oo{} molecular beam (left).
Here, the target surface was moved away from the line of sight of the molecular beam.
Subsequently, we positioned the target surface in line with the molecular beam.
This adjustment caused a pressure change, attributed to an increase in hydrogen gas pressure, as confirmed by the mass spectrometer. 
The surface was still clean as confirmed by AES measurements.
With the target surface in place, we turned on the oxygen beam and monitored the background pressure in the UHV chamber using the ion gauge. 
By subtracting the flux of molecules not absorbed from the total flux, we calculated the amount of oxygen adhered to the surface.

The dose of adsorbed oxygen molecules, estimated using the uncalibrated ion gauge, was equivalent to 0.285 ML, equating to an atomic coverage of 0.57 ML on the Cu(110) surface (see figure \ref{FIG_03}, right). 
Since it is well-known that the saturation coverage should be 0.5 ML, we conclude that our ion gauge overestimates the pressure by 14\%. 
This calibration along with the gas-dependent sensitivity factors was used throughout our analysis for incident flux estimation.

As an additional confirmation, we also determined the initial sticking probability of \oo{} to be 0.36, which is consistent with that reported previously. \cite{nesbitt_O2_Cu110_1991}.

\section {SI-4: Molecular beam profile at the Cu(110) surface}

\begin{figure}[H]
\includegraphics[width=.8\linewidth]{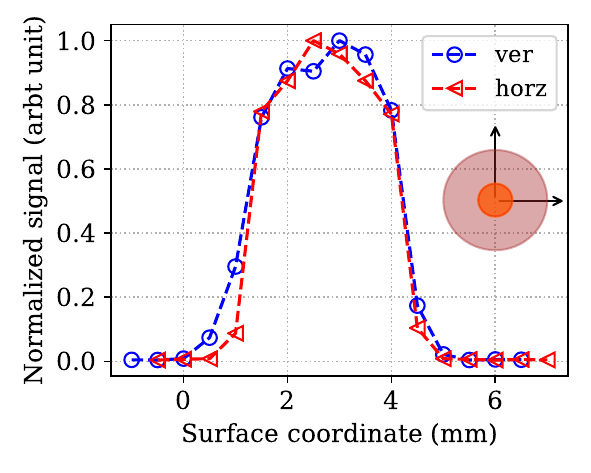}
\caption{Beam profile of the \coo{} beam (1.40 eV) incident on the Cu(110) surface, measured by mapping the spatial profile of the O-atom coverage. 
The blue and red curve represents the beam profile along the vertical and horizontal direction, respectively. 
}
\label{FIG_04}
\end{figure}

Another important element to determine the absolute incident dose accurately is the shape of the molecular beam on the target surface. 
We accomplished this by measuring a spatial profile of the chemisorbed oxygen resulting from dissociative \coo{} at the saturation coverage limit on the Cu(110) surface. 
The molecular beam profile was measured along both the vertical and horizontal axes (see figure \ref{FIG_04}).
We find that the cross-sectional area of our incident molecular beam corresponds to a circle with a diameter of 2.9 mm.

\newpage

\section {SI-5: Estimating the flux of the incident beam}

Following the ion gauge calibration and determination of the beam profile on the target surface, we proceeded to establish a correlation between the ion gauge and the mass spectrometer measurements. 
In the case of a molecular beam mixture (\coo{} + \hh{}), relying solely on the total pressure change measured using an ion gauge is not sufficient to calculate the incident dose of \coo{}. 
To address this, we measured the individual partial pressures of \hh{} and \coo{} using a mass spectrometer at the time of dosing. 
The mass spectrometer itself was calibrated against the ion gauge across a pressure range of $1 \times 10^{-9}$ to $8 \times 10^{-9}$ mbar.

\begin{figure}[H]
\includegraphics[width=.9\linewidth]{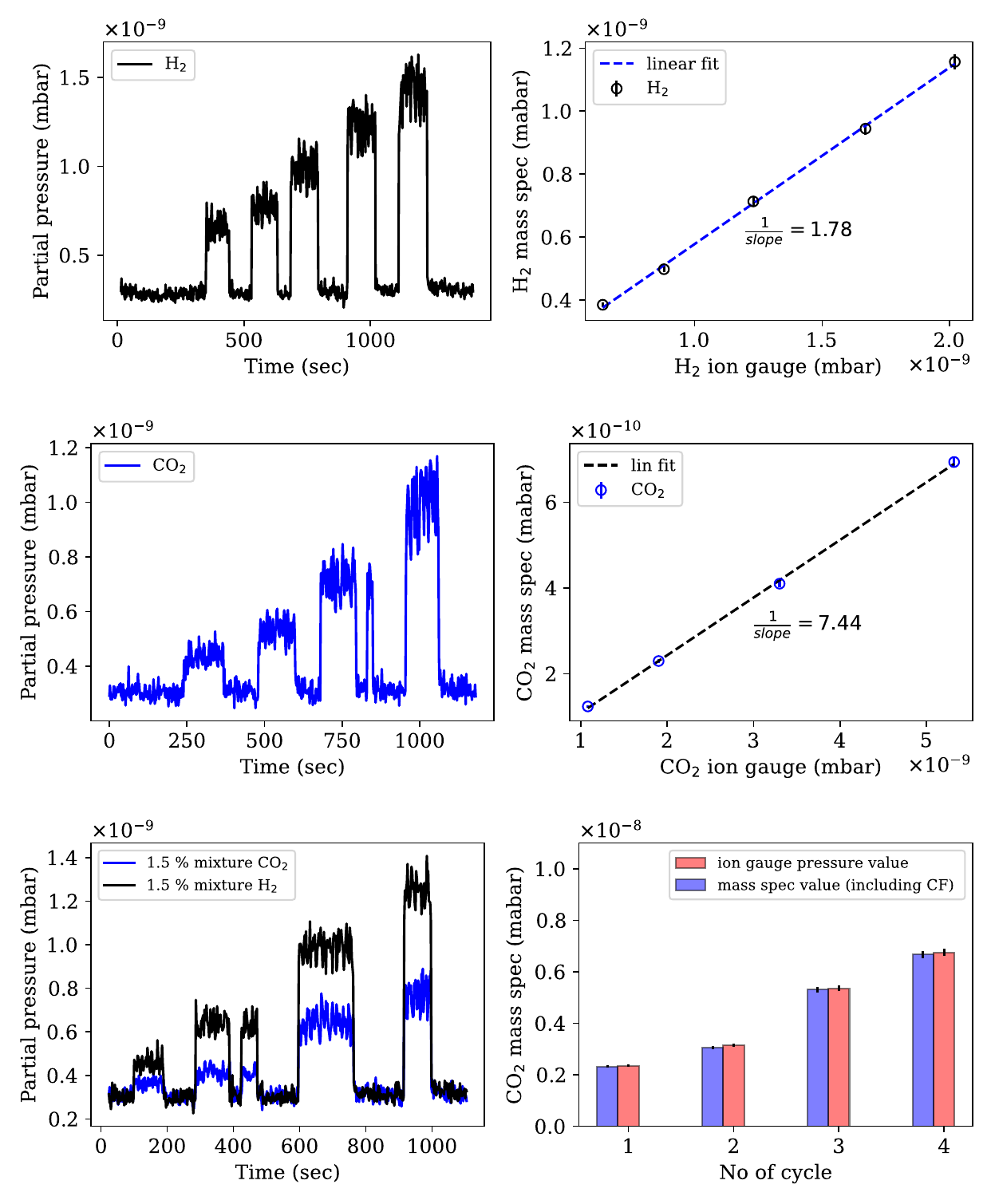}
\caption{(upper left) The dose of a pure \hh{} beam measured using the mass spectrometer for various working pressure ranges. 
The upper right panel demonstrates the relationship between the measured ion gauge pressure and the corresponding measurement using the mass spectrometer.
A linear fit to this data yielded a calibration factor of 1.78 for \hh{}.
The middle panel shows the same for \coo{} resulting in a calibration factor of 7.44. 
To further validate this procedure for a gas mixture, individual partial pressures were measured in the mass spectrometer and corrected using the calculated calibration factors. 
The corrected total pressure obtained was in  excellent agreement with the total pressure measured by the ion gauge. 
The uncertainty in these measurements remained well within 5\%.}
\label{fig:FIG_05}
\end{figure}

Measurements using pure \hh{} and \coo{} individually, revealed that the mass spectrometer underestimated the partial pressures of \hh{} and \coo{} by a factor of 1.78 (see Figure \ref{fig:FIG_05}, upper panel) and 7.44 (see Figure \ref{fig:FIG_05}, middle panel), respectively.
It should be noted that at the moment, these factors do not include the gas-dependent sensitivity factors of the ion gauge itself (they are included separately later).
To further confirm if these calibration factors hold true for a gas mixture, we measured the change in total pressure using the ion gauge upon turning on a molecular beam of 1.5\% \coo{} in \hh{} and simultaneously the respective partial pressures using the mass spectrometer. 
The sum of the individual partial pressure changes, adjusted using the estimated calibration factors above, was found to be in agreement with the total pressure change measured with the ion gauge ion (see Figure \ref{fig:FIG_05}, lower panel).
The estimated uncertainty for these measurements remained below 5\%.

\subsection*{Incident \coo{} dose estimation}

The true partial pressure change of \coo{} upon turning on the incident molecular beam, measured using the mass spectrometer was calculated as follows:
 
\begin{equation}
 \text{True pressure change of \coo{}: 
   }  
   \Delta P_{true} = \frac{CF_{\coo{}}\times \Delta P_{obs}}{IG_{SF_{\coo{}}}}
\end{equation}
where $CF_{\coo{}}$, $P_{obs}$ and $IG_{SF_{\coo{}}}$ stand for calibration factor for \coo{} (7.44), pressure change during dosing and ionization gauge calibration factor for \coo{} (1.42), respectively.

 \begin{equation}
 \text{Number of molecules/sec:    }   N_{\coo{}} = \frac{\Delta P_{true} \times \text{pumping speed} \times N_a }{R_g \times T}    
 \end{equation}

\begin{equation}
 \text{Number of molecules/sec/cm$^2$, Flux:    } \phi_{\coo{}} = \frac{N_{\coo{}}}{\pi \times r^2}
 \end{equation}

 \begin{equation}
  \text{Dose in ML per sec}= 
  \frac{\phi_{\coo{}}}{1.08 \times 10^{15} \text{(atoms/cm$^2$/ML)}}    
 \end{equation}

\section {SI-6: Kinetic model for estimating surface coverage caused by non-dissociative physisorption and dissociative chemisorption }

To understand the reasons behind the conclusion drawn in the previous work \cite{burghaus_CO2_Cu110_2006} that \coo{} dissociation was absent, we constructed a kinetic model based on simple first-order kinetics. At a specific incident energy of \coo{}, molecules undergo both trapping and dissociative adsorption. Their experimental conditions involved low surface temperatures (below 90 K), leading to the occupation of sites by both \coo{}, CO, and O (both resulting from dissociation). 
A single \coo{} molecule is assumed to occupy one surface site, while a combination of CO and O molecules can also block one site each \cite{yimin_APXPS_DFT_2020}. 

The maximum saturation coverage for physisorbed \coo{} can reach up to 1 ML, whereas for CO and O, it is limited to 0.25 ML each. 
Having this information, we formulated rate equations for each elementary process as shown below:
\begin{align}
\ce{CO_2{_{(g)}} + \text{*}} & \ce{->[k_{1}]  CO_2_${phy}$}
\end{align} 

\begin{align}
\ce{CO_2{_{(g)}} + 2\text{*}} & \ce{->[k_{2}] CO_${phy}$ + O_${chem}$}
\end{align} \vspace{.6 cm}

Here, $k_{1}$ is the non-dissociative initial sticking probability for \coo{}, $k_{2}$ is the dissociative sticking probability of \coo{} and * denotes an available site on Cu(110) surface.
The rate equations for overall surface coverage at any given time t ($\Theta(t)$) can be expressed as follows:

\begin{align}
\frac{d[\Theta(t)]}{dt} = (\ k_{1} +\ k_{2})(1-{\Theta(t)})[\rm{CO_2{_{(g)}}}]
\label{kinetics}
\end{align} 

\begin{align}
\Theta(t) = 1- e^{-(k1+k2)t[\rm{CO_2{_{(g)}}}]}
\end{align} 

\begin{align}
\frac{d[{{\rm{CO}}_2{_{phy}}}]}{dt} = \ k_{1}(1-{\Theta(t)})\rm[{CO_2{_{(g)}}}] 
\label{kinetics}
\end{align}  

\begin{align}
\frac{d\rm[{{\rm{CO}}{_{phy}}}]}{dt} = \ k_{2}(1-{\Theta(t)})\rm[{CO_2{_{(g)}}}]
\label{kinetics}
\end{align} 

\begin{align}
\frac{d[{\rm{O}}_{chem}]}{dt} = \ k_{2}(1-{\Theta(t)})\rm[{CO_2{_{(g)}}}] 
\label{kinetics}
\end{align} 
Upon substituting the value of $\Theta(t)$ into equations 9-11, and subsequently solving the resulting differential equation while applying appropriate boundary conditions (\coo{} saturation = 1 ML, CO, O-atom saturation coverage = 0.25 ML each and starting with a clean surface), we obtain:

\begin{align}
{[{\rm{CO}}_{2}{_{phy}}(t)]} = \frac{k_{1}}{(k_{1}+\ k_{2})} -\frac{\ k_{1}}{(\ k_{1}+\ k_{2})}e^{-(\ k_{1}+\ k_{2}){\rm{[\coo{}}_{(g)}]}t}  
\label{kinetics}
\end{align} 

\begin{align}
{[{\rm{CO}}_{phy}(t)]} = 0.25\frac{\ k_{2}}{(\ k_{1}+\ k_{2})} -0.25\frac{\ k_{2}}{(\ k_{1}+\ k_{2})}e^{-(\ k_{1}+\ k_{2}){\rm{[\coo{}}_{(g)}]}t} 
\label{kinetics}
\end{align} 

\begin{align}
{[{\rm{O}}_{chem}(t)]} = 0.25\frac{\ k_{2}}{(\ k_{1}+\ k_{2})} -0.25\frac{\ k_{2}}{(\ k_{1}+\ k_{2})}e^{-(\ k_{1}+\ k_{2}){\rm{[\coo{}}_{(g)}]}t} 
\label{kinetics}
\end{align} 

The derived solution yields the time-dependent coverage buildup of physisorbed \coo{}$_{phy}$, CO$_{phy}$, and chemisorbed O$_{chem}$ on the Cu(110) surface. 
With knowledge of the rate constants, $k_1$, and $k_2$, we have estimated the final coverage of \coo$_{phy}$, CO$_{phy}$, and chemisorbed O$_{chem}$ on the surface. 

\newpage
\section {SI-7: Influence of defects on \so{}} 

\begin{figure}[H]
\includegraphics[width=1\linewidth]{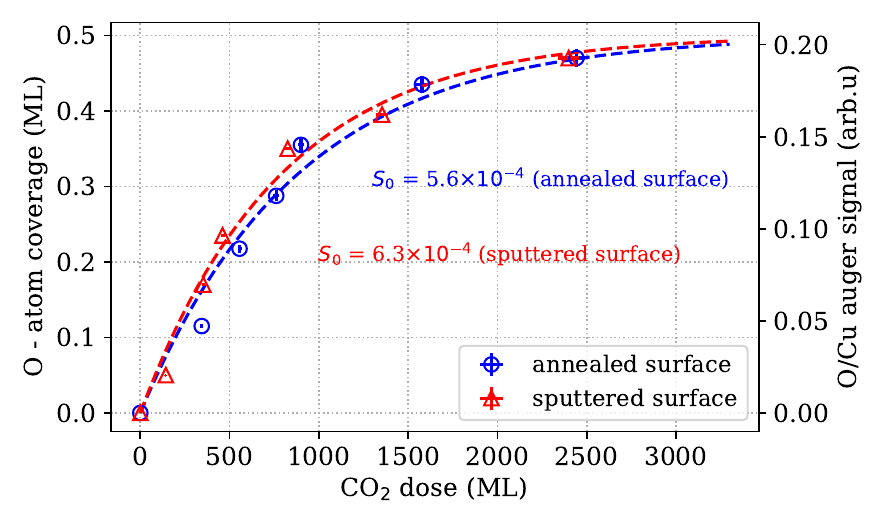}
\centering
\caption{Dissociative adsorption kinetics of CO$_2$ on both clean and Ar ion-sputtered Cu(110) surfaces. 
Under our measurement conditions, we were unable to observe a significant change in the \so{} values.}
\label{fig:fig_clean_defect}
\end{figure}

We investigated the influence of surface defects on the \so{} of \coo{} dissociation on Cu(110) by comparing measurements on a clean and annealed vs clean and Ar ion-sputtered (without annealing) surface (see figure \ref{fig:fig_clean_defect}.
Here, a \coo{} beam with \ei{} = 1.15 eV was used.
The clean Cu(110) surface was sputtered for 10 minutes at $T_s$ = 475 K, with a surface current of 0.5 $\mu$A.
We estimate that the dose of Ar ions incident on the surface to be 1.2 ML. 
Under these conditions, we were unable to observe any significant change beyond the experimental uncertainties in the \so{} values.
Further, systematic studies with larger sputtering times will be needed to understand this point better.

\section {SI-8:  Discussion on possible contamination in our incident beams}

In this study, we concluded the dissociation of \coo{} on Cu(110) by measuring the dissociative chemisorbed oxygen signal using auger electron spectroscopic methods. 
However, our measurements could potentially be affected by the presence of oxygen-containing gas molecules such as CO, H$_2$O, \oo{} in the molecular beam, leading to an accumulation of oxygen on the Cu(110) surface.

Presence of any \oo{} contamination was ruled out by the fact that dosing with pure \coo{} beam did not result in any measurable O-atom coverage.
Further, we monitored the signals of \coo{}, \hh{}, CO, and \oo{} using a mass spectrometer during  dosing. 
Notably, we did not observe any discernible change in the partial pressures for gases other than \coo{} and \hh{}, which are components of the mixture. 
Our observations allow us to largely rule out the presence of any contamination within the sensitivity of our mass spectrometer.
Furthermore, if there were any significant oxygen coverage resulting from the presence of CO contamination in our beam (such as CO dissociation), we would expect to observe carbon peaks too. 
However, no such peaks were detected in our measurements, thereby ruling out CO contamination.

Another potential source of contamination is that of H$_2$O. 
This can potentially arise from the reverse water gas shift (RWGS) reaction occurring within the gas mixture. 
If such a reaction were to occur, both \hh{}O and CO would be formed, and we would anticipate a measurable change in the CO signal in the mass spectrometer. However, no significant changes in the \coo{} to CO partial pressure ratio were observed during dosing for all mixtures (within 3\%).
Even if we assume that the \hh{}O formed is below our detection sensitivity, we estimate it to be of the order of maximum 3\% of the \coo{} fraction. 
The \ei{} range for such dilute mixtures of \hh{}O will only span 0.05 eV. 
Such minute energy changes will not lead to energy-dependent initial sticking probabilities, as seen in our measurements.
Based on this analysis, we rule out any role of \hh{}O dissociation on the results presented.

\bibliography{bibliography}